# Noise-based logic: Binary, multi-valued, or fuzzy, with optional superposition of logic states

Laszlo B. Kish[a]

Texas A&M University, Department of Electrical and Computer Engineering, College Station, TX 77843-3128, USA; email: Laszlo.Kish@ece.tamu.edu

**Abstract**. A new type of deterministic (non-probabilistic) computer logic system inspired by the stochasticity of brain signals is shown. The distinct values are represented by independent stochastic processes: independent voltage (or current) noises. The orthogonality of these processes provides a natural way to construct binary or multi-valued logic circuitry with arbitrary number $N$ of logic values by using analog circuitry. Moreover, the logic values on a single wire can be made a (weighted) superposition of the $N$ distinct logic values. Fuzzy logic is also naturally represented by a two-component superposition within the binary case ($N=2$). Error propagation and accumulation are suppressed. Other relevant advantages are reduced energy dissipation and leakage current problems, and robustness against circuit noise and background noises such as $1/f$, Johnson, shot and crosstalk noise. Variability problems are also nonexistent because the logic value is an AC signal. A similar logic system can be built with orthogonal sinusoidal signals (different frequency or orthogonal phase) however that has an extra $1/N$ type slowdown compared to the noise-based logic system with increasing number of $N$ furthermore it is less robust against time delay effects than the noise-based counterpart.

---

[a] Until 1999: L.B. Kiss

## 1. Introduction

Recently, new, non-conventional ways of stealth communications [1] and unconditionally secure communications [2] were introduced and the last one was successfully demonstrated [3]. These methods used electronic thermal noise and its enhanced versions as information carrier. Inspired by this success, a new type of computation was envisioned where the thermal noise and its statistical properties would be the information carrier [4]. For a short review of the results in [1-4] and that of other non-conventional informatics, see [5]. In [4] no concrete solution could be proposed for thermal noise-driven computers, however a lower limit of the energy dissipation in the order of $kT$/bit was given, where $k$ is the Boltzmann constant and $T$ is the temperature.

One reason to explore such an unconventional way of computing was the fact that neural signals are stochastic processes, thus the brain is also using noise and its statistical properties for information processing. Another set of reasons was the numerous problems with current microprocessors and the miniaturization to follow Moore's law [6]. Today's computer logic circuitry is a system of coupled DC amplifier stages and this situation represents enhanced vulnerability against *variability* [7] of fabrication parameters such as threshold voltage-inaccuracies in CMOS. Thin oxides imply great power dissipation due to leakage currents [7]. Thermal noise, its error generation and the power dissipation need of reducing these problems are another issue [9-11]. Famous initiatives such as quantum computers or reversible computing are unable to help in these regards [12,13]. Tough there are various interesting proposals [14-19] to "live with noise" and to improve these conditions, major breakthroughs are still needed to secure future evolution of performance.

As a next step forward in the direction shown in papers [1-4], in this Letter, we introduce a new, non-conventional logic initiative where the logic values are carried by independent noise processes. At the end of the paper, we briefly show and compare another possibility, when the logic values are carried by sinusoidal signals with different frequencies or orthogonal initial phase, however the noise-based logic seems more feasible.

## 2. Motives, and introducing noise-based logic

A general problem with today's binary logic is that, the non-zero DC voltage levels representing the basic logic values (0,1) ("Low", "High") can be considered vectors with different length but in the same direction. These vectors are not *orthogonal* to each other, or to background noises, transients/spikes, such as ground-plane EMI and cross-talk pulses because these all can be considered as parallel vectors. Thus, the shrinking noise margin with miniaturization implies a progressively increasing rate of dynamical errors.

In the *noise-based logic* scheme proposed in this paper, the logic values are represented by *independent stochastic processes* (electronic noises) of *zero mean*. A noise-based logic gate typically contains analog circuitry, such as linear amplifiers, multipliers, filters

(especially time averaging units) and analog switches. This situation has several immediate advantages:

**i.** The different basic logic values are orthogonal not only to each other but also to any transients/spikes or any background noise including thermal noise or circuit noise, such as *1/f*, *shot*, *generation-recombination*, etc, processes.

**ii.** Due to the zero mean of the stochastic processes, the logic values are AC signals and AC coupling can make it sure that the variability-related vulnerabilities are strongly reduced.

**iii.** Switching errors. The noise-based logic values have a reduced impact because these switching errors are also independent stochastic processes thus also orthogonal to the logic values. This property implies a reduced energy need to run the switches. Moreover, these errors do not propagate and accumulate in the proposed noise-based logic system.

**iv.** Due to the orthogonality and AC aspects (points **i** and **ii**), a noise carrying a given logic value on the data bus can have much smaller effective value than the power supply voltage of the chip. This property is also very different from today's digital circuitry and it can offer another way of reducing the energy consumption.

**3. Mathematical definition of noise-based logic**

Mathematically, noise-based logic is based on an orthogonal basis of time functions $V_i(t)$ ($i = 1...N$) similarly to Fourier analysis/synthesis, however here the base functions are independent electronic noises: different realizations of the same or different Gaussian stochastic processes with zero mean. Each base function represents a different logic value and $N$ is the total number of logic base values. For example, in the binary case ($N = 2$):

$$\langle L^2(t) \rangle = 1 \quad , \quad \langle H^2(t) \rangle = 1 \quad \text{and} \quad \langle H(t)L(t) \rangle = 0 \quad , \tag{1}$$

where $\langle \ \rangle$ represents time average; the $L(t)$ and $H(t)$ noise processes represent the "Low" and "High" logic values, respectively; and for simplicity we supposed that the *RMS* value of the noises is 1. Note, that unlike in digital systems, the "Low" state has the same signal amplitude as the "High" state. Generally, for arbitrary number of logic values:

$$\langle V_i(t)V_j(t) \rangle = \delta_{i,j} \tag{2}$$

where $\delta_{i,j}$ is the Kronecker symbol (for $i = j$, $\delta_{i,j} = 1$, otherwise $\delta_{i,j} = 0$). Due to equation 1 or 2, the $L(t)$, $H(t)$ or $V_i(t)$ processes can be represented by orthogonal unit vectors in multidimensional space, thus we can use the term *logic base vectors* for *logic base values* and talk about *N*-dimensional logic space and *logic state vectors* in it.

Generally, a logic state vector is the *weighted* superposition of logic base vectors:

$$X(t) = \sum_{i=1}^{N} a_i V_i(t) \qquad (3)$$

If only one of the $a_i$ coefficients differs from zero, we have a "clean" multivalue logic otherwise we have a superposition which can be either a discrete superposition with discrete coefficient (for example $a_i = 0$ or $a_i = 1$), or continuum, fuzzy logic (for example $0 \leq a_i \leq 1$).

## 4. An example of coefficients: binary noise-based logic scheme, and its fuzzy version

When $N = 2$, see above, and the coefficients $a_1$ and $a_2$ are either 0 or 1 and $a_1 \neq a_2$, we have a standard binary noise-based logic, see Figure 1, were the vector representation of noise-based logic with the two base values is shown. When the logic is binary, only single base vectors are used. If a superposition $X(t) = a_L L(t) + a_H H(t)$ is used, it is a fuzzy logic with continuum values along the unit circle if we suppose that the normalization condition, $a_L^2 + a_H^2 = 1$, holds. Without normalization, the vector can point anywhere in that quarter plane ($x, y \geq 0$).

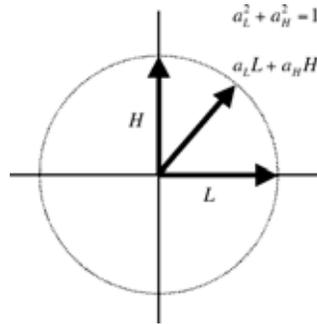

**Figure 1.** Vector representation of noise-based logic with two base values. When only single base vectors are used, the logic is binary. If a superposition is used, the logic is fuzzy with continuum values along the unit circle.

## 5. General classes of noise-based logic

In general, the noise-based logic gates to be described below can work in the following modes, depending on their infrastructure. Note, superposition means the superposition of logic states on a single wire.

**1.** *Discrete state vectors* in finite, $N$-dimensional logic space, *without superposition*; see the $N = 2$ case in Figure 1. Even though the system works with the noise-based logic values and with building elements of analog circuitry (see below), the practical accuracy of this system is identical to that of a *digital circuitry*. However in several practical aspects it is superior to the normal two-value (binary) logic, even when the noise based-logic is used in the binary mode, see points **i-iv** above.

**2.** *Discrete state vectors* in finite, *N*-dimensional logic space, *with superposition* (with discrete coefficients). Same as at point **1**, however with superposition of the discrete logic values with *discrete, digital-accuracy coefficients*. In the binary case ($N = 2$), such a superposition represents the usual fuzzy logic, see the superposition vector in Figure 1. The simplest situation of the $N > 2$ case is when the coefficients $a_i$ can only be 1 or 0 (a base element is "on" or "off" in the superposition). In the more general cases, to provide high resolution for the coefficients with *digital accuracy*, even the value of these coefficients must be represented by independent noise processes. Then, in the case of many discrete values, it is appropriate to use a "digital fashion" for the coefficient values and to talk about how many bit resolution is needed for these coefficients. The weighting coefficients $a_i$ will be replaced by $b_{i,j} c_j(t)$, where $c_j(t)$ is the $j$-th bit of the coefficient values; $b_{i,j}$ is the digitized bit coefficient ($b_{i,j} = 0$ or 1); and the index "$i$" is related to the base value $V_i$. If the required digital accuracy of the coefficients is $M$ bits and the logic base is $N$-dimensional, then:

$$X(t) = \sum_{i=1}^{N} \sum_{j=1}^{M} b_{i,j} c_j(t) V_i(t) \qquad (4)$$

**3.** *Continuum state vectors* in finite, *N*-dimensional logic space, *superposition with continuum coefficient* values. In this case, Equation 3 is relevant however with continuum values of the constants $a_i$. Its disadvantage is that the accuracy of this system is much lower because the coefficients must be analog numbers; so this is only a kind of "multidimensional" analog logic.

**4.** *Logic hyperspace* with discrete or continuum vectors. It is based on the fact that the product of two independent noise processes will be a third noise process which is independent from all the former base vectors therefore it serves as a new base vector in a new dimension, the *logic hyperspace*:

$$\text{If} \quad i \neq k \quad \text{and} \quad H_{i,k}(t) \equiv V_i(t) V_k(t) \quad \text{then for all} \quad n = 1...N, \quad \langle H_{i,k}(t) V_n \rangle = 0 \qquad (5)$$

Similar operation can be done with the hyperspace vectors to gain extra hyper space:

$$\text{If} \quad L_{i,k,l,m}(t) \equiv H_{i,k}(t) H_{l,m}(t) \quad \text{then} \quad \langle L_{i,k,l,m}(t) V_n \rangle = 0, \quad \langle L_{i,k,l,m}(t) H_{p,q} \rangle = 0 \qquad (6)$$

provided $i \neq k \neq l \neq m$.

Thus the number of dimensions of an originally *N*-dimensional logic space can arbitrarily be expanded by specific logic gates, which make the product of two independent vectors in the original space, or that of the hyperspace. The logic state vectors in the hyperspace cannot be interpreted as the linear combinations of the original base vectors and they can be utilized for non-conventional logical operations.

## 6. On the conceptual building elements of noise-based gates

From Equations 1-4 it is obvious that the basic building elements of noise based-logic (out of the noise generators which can be simply resistors or transistors) are the same as that of analog computers: linear amplifiers; analog multipliers; adders; linear filters, especially time average units which are low-pass filters; analog switches; etc.

At the input of the logic gate, the input state must be evaluated. This is typically done by multiplication with the relevant reference noises and followed by a time averaging:

$\langle X_1(t)V_j(t) \rangle$, $\langle X_2(t)V_j(t) \rangle$, $\langle X_1(t)V_k(t) \rangle$, $\langle X_2(t)V_k(t) \rangle$, etc.

The result will be a zero or nonzero value (DC voltage), which, with the other similar terms, will drive the internal logic units: typically controlling analog switches connected to reference noise signals, see below. Mathematically, this can be represented by multiplying the DC voltage (or its scaled/amplified version) with the relevant reference noise voltage and averaging the result, making the cross-correlation, see below.

Before showing the binary gate examples, is important to address the practical situation, which means finite-time averages that imply statistical fluctuations of these averages. During the rest of the paper, we suppose a finite time average with $\tau$ duration, even if we write $\langle \; \rangle$ instead of $\langle \; \rangle_\tau$. Furthermore, to define discrete logic values with *digital accuracy*, we must introduce upper and lower saturation operations represented by under/over bars as follows:

The operation $\overline{\underline{A}}$ means that if $A \geq U_{HT}$ then $\overline{\underline{A}} = 1$, and if $A \leq U_{LT}$ then $\overline{\underline{A}} = 0$ (7)

The saturation operations are important also to deal with the limited accuracy of analog circuitry, including the non-idealities of multipliers. Thus simple circuits such as classical high-frequency mixer arrangements can also be considered for multipliers. These saturation operations are naturally present in CMOS amplifier stages due to the threshold voltage of the MOS transistors. They suppress error propagation and accumulation in the noise-based logic, which is a unique property, see below. Out of these, they often play the role of a normalization-type of operation; see several cases below, when the reference noise is multiplied by a sum that sometimes can have greater value than 1.

The complementary operation of (7) is not essential but practically useful:

The operation $\overline{\underline{A}}^{\otimes}$ means that if $A \geq U_{HT}$ then $\overline{\underline{A}}^{\otimes} = 0$, and if $A \leq U_{LT}$ then $\overline{\underline{A}}^{\otimes} = 1$ (8)

The conceptual analog circuit elements that we will use for the realization of the logic gate examples are shown in Figure 2.

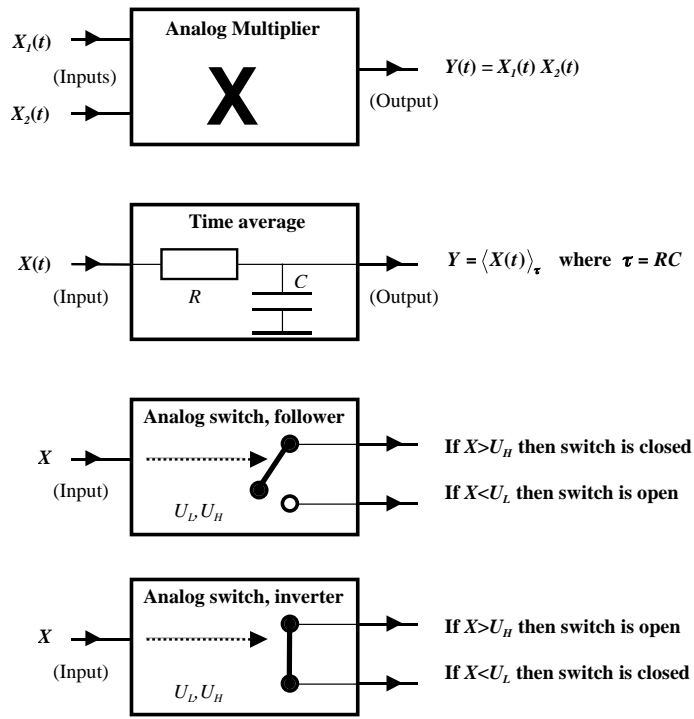

**Figure 2.** The analog circuit elements that can be used to realize the gates below. There are some other trivial building elements, such as linear adders and amplifiers, not shown here.

The time average circuit can be the standard *RC* low-pass filter which provides an averaging time $\tau = RC$. The *follower* analog switch is closed when the input voltage is greater than $U_H$ and opened when it is less than $U_L$ (these threshold values are the same as used before). The *inverter* analog switch behaves in the opposite fashion.

### 7. Example: binary noise-based logic functions

We give the complete mathematical foundation of binary noise-based logic with digital accuracy by showing its basic logic gates. Because arbitrary binary logic functions can be realized by OR, AND, and INVERTER gates, the foundations of the binary ($N = 2$) noise-based logic are represented by the noise-based version of these gates. The input variables are represented by $X_i$, the output by $Y$ and the logic base vectors (noises) are defined by Eq. 1, $H(t)$ for logic "High" and $L(t)$ for logic "Low".

Generally, noise-based logic gates have one logic output, several reference inputs (where the logic base vector noises are connected), and one or more logic inputs, see for example the binary INVERTER gate in Figure 3.

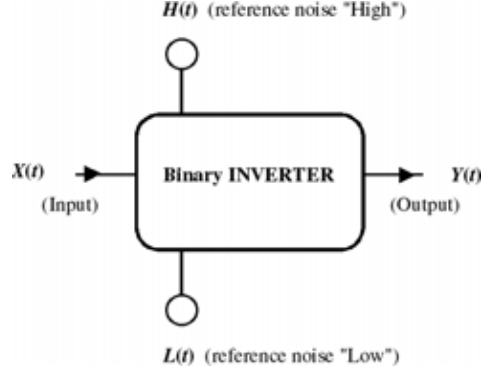

**Figure 3.** The logic and reference inputs and the output of the binary INVERTING gate.

The noise-based logic functions of the basic binary gates are as follows.

Binary INVERTER gate (Table 1):

$$Y(t) = \overline{\langle X(t)H(t)\rangle}L(t) + \overline{\langle X(t)L(t)\rangle}H(t) \quad , \tag{9}$$

or the more practical version with less multiplier elements:

$$Y(t) = \overline{\langle X(t)H(t)\rangle}L(t) + \overline{\langle X(t)H(t)\rangle}^{\otimes}H(t) \quad . \tag{10}$$

| $X(t)$ | $Y(t)$ |
|---|---|
| $H(t)$ | $L(t)$ |
| $L(t)$ | $H(t)$ |

**Table 1**. Binary INVERTER gate truth table.

Binary OR gate (Table 2):

$$Y(t) = \overline{\left[\langle X_1(t)H(t)\rangle + \langle X_2(t)H(t)\rangle\right]}H(t) + \overline{\left[\langle X_1(t)L(t)\rangle\langle X_2(t)L(t)\rangle\right]}L(t) \tag{11}$$

or the more practical version with less multiplier elements:

$$Y(t) = \overline{\left[\langle X_1(t)H(t)\rangle + \langle X_2(t)H(t)\rangle\right]}H(t) + \overline{\left[\langle X_1(t)H(t)\rangle + \langle X_2(t)H(t)\rangle\right]}^{\otimes}L(t) \tag{12}$$

| $X_1(t)$ | $X_2(t)$ | $Y(t)$ |
|---|---|---|
| $H(t)$ | $H(t)$ | $H(t)$ |
| $L(t)$ | $H(t)$ | $H(t)$ |
| $H(t)$ | $L(t)$ | $H(t)$ |
| $L(t)$ | $L(t)$ | $L(t)$ |

**Table 2**. Binary OR gate truth table.

Binary AND gate (Table 3):

$$Y(t) = \overline{\left[\langle X_1(t)H(t)\rangle\langle X_2(t)H(t)\rangle\right]}H(t) + \overline{\left[\langle X_1(t)L(t)\rangle + \langle X_2(t)L(t)\rangle\right]}L(t) \quad , \quad (13)$$

or the more practical version with less multiplier elements:

$$Y(t) = \overline{\left[\langle X_1(t)H(t)\rangle\langle X_2(t)H(t)\rangle\right]}H(t) + \overline{\left[\langle X_1(t)H(t)\rangle\langle X_2(t)H(t)\rangle\right]^{\otimes}}L(t) \quad (14)$$

| $X_1(t)$ | $X_2(t)$ | $Y(t)$ |
|---|---|---|
| $H(t)$ | $H(t)$ | $H(t)$ |
| $L(t)$ | $H(t)$ | $L(t)$ |
| $H(t)$ | $L(t)$ | $L(t)$ |
| $L(t)$ | $L(t)$ | $L(t)$ |

**Table 3**. Binary AND gate truth table.

Equations 1, 7-8, and 9-14 provide the foundations of binary noise-based logic.

Binary XOR (with possible multi-value input state, Table 4). Though it is not necessary, for its versatility and easy realization, we show the XOR gate, too. The output is "High" if the two logic inputs see different values, otherwise the output is "Low". Note, the inputs can be driven by values N>2, too, and the gate still will function as XOR and it shows if the inputs are different or identical:

$$Y(t) = \overline{\langle X_1(t)X_2(t)\rangle^{\otimes}}H(t) + \overline{\langle X_1(t)X_2(t)\rangle}L(t) \quad (15)$$

| $X_1(t)$ | $X_2(t)$ | $Y(t)$ |
|---|---|---|
| $H(t)$ | $H(t)$ | $L(t)$ |
| $L(t)$ | $H(t)$ | $H(t)$ |
| $H(t)$ | $L(t)$ | $H(t)$ |
| $L(t)$ | $L(t)$ | $L(t)$ |

**Table 4**. Binary XOR gate truth table.

Finally we show another multi-value application, a 4-input XOR gate with binary output and arbitrary number $N \geq 4$ of input logic values. The output is "True" whenever the four input values are different, otherwise it is "False". Compared to the *infinite number* of different input possibilities, the solution is very simple: it needs 6 multipliers, one time average unit, one follower switch and one inverter switch (compare Equation 15 and its simplified realization with switches):

$$Y(t) = \overline{\langle X_1(t)X_2(t) + X_1(t)X_3(t) + X_1(t)X_4(t) + X_2(t)X_3(t) + X_2(t)X_4(t) + X_3(t)X_4(t)\rangle} L(t) +$$

(16)

$$+ \overline{\langle X_1(t)X_2(t) + X_1(t)X_3(t) + X_1(t)X_4(t) + X_2(t)X_3(t) + X_2(t)X_4(t) + X_3(t)X_4(t)\rangle}^{\otimes} H(t)$$

## 8. A circuit example of noise-based logic gates: XOR and INVERTER gates

Equation 15 is very simple, just like its circuit realization, which we show as an example, how the hardware scheme of noise-based logic may look like. Figure 4 shows the binary XOR circuit, which is identical with the INVERTER circuit, except that the input of the INVERTER is the first input and the second input is connected to reference $H(t)$.

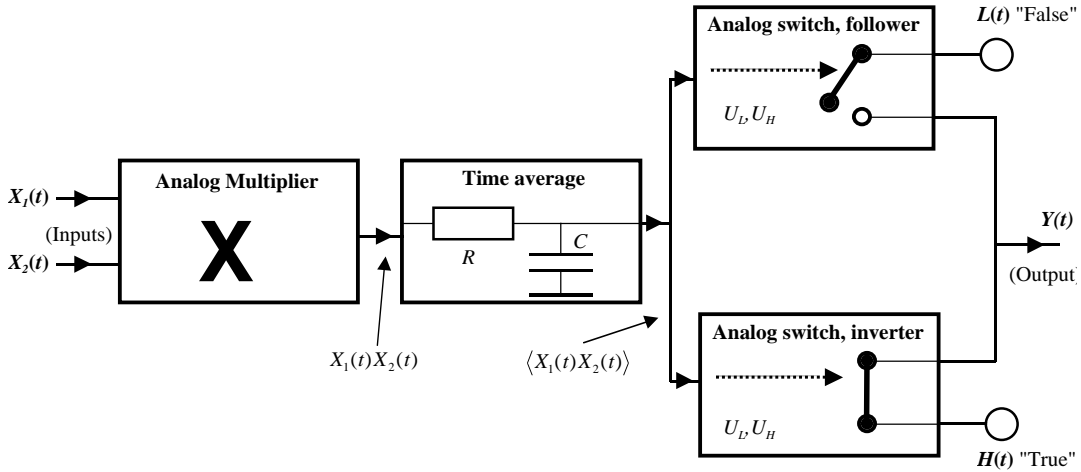

**Figure 4.** Simple, noise based XOR circuit (binary or multi-value-input). If the two inputs (or arbitrary number of logic values) are equal, the output is *False*, otherwise it is *True*. If the second input is connected to reference $H(t)$, we get an INVERTER gate.

It is easy to see why using equations 10, 12, 14 and 15, with the complementary operation and less multipliers is much more economical than using equations 9, 11 and 13 instead. The single multiplier output after time averaging can drive a complementary pair of analog switches (mathematically an alternating switch however with transistors it will always be two switches). Though the equations contain several multiplications, at the circuit level a single multiplier is enough for any of these gates.

**9. A few observations and illustrative estimations**

Final and detailed assessment about energy dissipation, speed and errors in noise based logic can only be done about a concrete hardware realization. In this section we are making some simple estimations at the general level, and these preliminary results are encouraging.

*Random errors in the switches do not accumulate and propagate.* In Figure 4, it is easy to see why the noise-based logic suppresses errors generated in the switches, even if these errors are generated inside the gate. If the two switches work somewhat erratically because the peak values of the noise sometimes crosses the threshold; these will be fast errors described by the Rice formula of level crossing [10]. Their lifetime is in the order of the correlation time $\tau_N \approx (2\pi B)$ of the noise where $B$ is the noise bandwidth of band-limited white noise. Then the output will not be a clean $H(t)$ or $L(t)$ but it will be a superposition of them. However this error will be cleared out in the next gate due to the threshold operations (Equations 7,8). For example, if we suppose that $U_H = U_L = 0.5$ and suppose an extraordinarily large switching error rate, such as 1% (cf. the $< 10^{-20}$ in normal digital logic [10]), the output will be in the superposition $0.99\,H(t) + 0.01\,L(t)$ or $0.99\,L(t) + 0.01\,H(t)$. In both cases, the values 0.99 and 0.01 are far away from the threshold thus the next gate will use the logic value with the coefficient 0.99. Thus the errors generated in the switches will not propagate or accumulate.

*The real errors* we must be concerned about are coming from the statistical inaccuracies at the output of the finite-time averaging because these errors are "slow" by matching the time scale of the logic operation. If the momentary averaging error reaches the threshold, it will produce a slow error at the output that will be picked up and will get through the averaging unit of the next logic gate. As an example, using the $U_H = U_L = 0.5$ condition above, let us suppose that the averaging output shows a temporary value 0.6 instead of the theoretically correct 0 value. Then the output of the circuit in Figure 4 will be in $L(t)$ state, instead of the correct $H(t)$, for duration of the order of the correlation time of the time-averaged noise.

For a quantitative estimation, let us suppose a white noise driven *RC* filter in Figure 4:

$$\tau_c = RC = (2\pi f_c)^{-1} \,, \tag{17}$$

where $\tau_c$ is the correlation time of filtered (time-averaged noise), $f_c$ is the cut-off frequency (-3dB) of this filter which is about the maximal practical frequency of logic gate operations. In the case of band-limited white noise at the input, the mean-square noise (error) $\langle U_{err}^2 \rangle$ at the output of the *RC* filter can be determined from the *RMS* amplitude $U_{inp}$ of the input noise and the cut-off frequency of the filter:

$$\langle U_{err}^2 \rangle = U_{inp}^2 \frac{f_c}{B} \tag{18}$$

We can estimate the mean frequency of such error events by using the Rice formula [6]. In [6,10] it was found by the Rice formula that a ratio of 12 between the noise margin (threshold) and the *RMS* background noise in classical digital circuits is safe to keep the error probability at sufficiently low level around $10^{-25}$ [10] for microprocessors in the foreseeable future. Many processors of today have much higher error rates however we shall keep to this idealistic requirement. Thus using the above assumptions:

$$12 \leq \frac{0.5}{\sqrt{\langle U_{err}^2 \rangle}} = \frac{0.5}{\sqrt{U_{inp}^2 \frac{f_c}{B}}} = \frac{0.5}{\sqrt{1 \frac{f_c}{B}}} = 0.5 \sqrt{\frac{B}{f_c}} \tag{19}$$

Thus,

$$\frac{B}{f_c} \geq 576 \tag{20}$$

Therefore for a noise bandwidth of 580 GHz, the speed of the logic would correspond to a 1 GHz classical logic speed with about $10^{-25}$ error probability during an operation. Such a high-speed circuitry needs a sub-50 nm technology [20] and if the small signal bandwidth is less than the above value, the speed of the noise-based logic operation will be proportionally less with this error limit.

*Dynamic power dissipation along the reference noise distribution network versus dynamic power dissipation of clock distribution networks of classical circuits.* One of the most significant sources of energy loss in microprocessors is the clock distribution network [21] because it goes everywhere, so its surface is large and it is fed by the fastest binary signal, the clock square waves of the processor. The reference noise network in a binary noise-based logic has similarly large surface; it has actually two such networks, one for the $H(t)$ another for the $L(t)$ reference. Capacitor currents at white noise drive will be $f^2$ noises thus, using the same *RMS* amplitude for the noise as for digital clock signals of frequency $f_c$, would yield a $\dfrac{2B}{f_c \sqrt{3}}$ times greater power dissipation than the digital clock. The factor 2 comes from the two noise networks and the $1/\sqrt{3}$ from the $f^2$ spectrum, respectively. However, the goal is to use as small reference noise signals as possible because the dissipation scales with the square of the amplitude. Using the condition in Eq. 20, a reference noise with amplitude 24 times less than the clock

amplitude (4% of that) would yield basically the same ($2/\sqrt{3}$ times greater) dissipation as that of the clock. Using noises with 1% of the clock amplitude would yield a 16 times less dissipation at the above conditions of 1GHz logic speed.

Note, in practical binary noise-based logic, it may often be more practical to generate the reference noises locally. A locally generated reference noise supplies the output of a given gate and the reference inputs of the other gates driven by that gate.

*No leakage current loss along the reference noise distribution network and strongly reduced leakage elsewhere.* In noise-based logic, because the leakage current scales exponentially with the voltage, it will be kept small and the leakage current dissipation will be negligible in the distribution network. Leakage current loss will also be strongly reduced in the logic gate circuitry because it can run by significant lower voltages then normally due to the low requirements about the switches (see below) due to the noise robustness against switching errors (see above).

*Strongly reduced voltage and energy need of driving the switches.* If a CMOS switch must connect and disconnect only small voltages, such as the reference noises (see above and Figure 4), it implies that the dynamic effective gate voltage controlling the switch can also be small. This is a further energy gain because the energy needed to control the gate capacitor *scales with the square* of this control voltage. This is an additional argument out of the robustness against switching errors. As a consequence, the supply voltage and energy need of the analog circuitry driving these switches can also be significantly reduced.

## 10. Interfacing with classical digital circuitry

It is also important to mention that the input/output interface of a binary noise-based logic to the external world of classical binary logic should naturally be built of the same elements: multipliers, time average units and switches. The only difference is that, in the input unit, the classical digital signal can directly drive the switches that are connected to the reference noise points just like in the circuitry in Figure 4. In the case of the output interface, however, the $L(t)$ references noise is replaced by zero (ground) voltage and the $H(t)$ by the positive supply voltage, respectively. Thus the output interface will provide the usual DC voltage levels. It is important to note that switching errors in the output interface will propagate to and accumulate in the classical digital units. Thus the output interface must run by the usual larger supply voltage and the energy saving arguments is not valid for this case.

## 11. A possible "sinusoidal relative" of noise-based logic

Finally, a word about possible similar logic spaces with sinusoidal time functions with different frequencies (or orthogonal phases) because they also form an orthogonal system. The main problems with them are the speed and/or errors. Even though at the first look such system may look faster because shorter averages maybe needed, as short as one period; at these short average times they are more prone to interferences (transients, noises, etc.) to which the noise-based logic is robust. They are more sensitive to phase shifts due to delay effects, too. Their weakest point can be seen at the multi-value logic. Then, to avoid crosstalk due to transient effects, the available small-signal bandwidth must be divided to at least $N$ non-overlapping sub-bands and locate the frequency carrier of logic values at the centers of such bands. Due to the amplitude modulation sidebands (left and right) produced by switching on/off these logic values, the maximally allowed averaging/calculating speed (clock frequency) will scale with the reciprocal of the width of such a sub-band $f_{c,\max} \approx \frac{B}{2N}$. This can be a significant slowdown compared to the noise-based logic which can always use the whole small-signal bandwidth.

## 12. About the noise generators

Noise sources are plenty at small sizes and there are various possibilities to generate reference noises. Perhaps the easiest way is to use the output noise of a MOS transistor, with possible weak amplification and impedance matching to drive the network. This noise would be $1/f$ below the GHz range and white above that. Because most of the noise power would be in the white range, our estimations in section 9 would hold. Noise sources of separate transistors are independent thus the orthogonality condition is easily satisfied.

## 13. Conclusions and some of the open problems

In conclusion, we have introduced a noise-based logic that, in its binary working mode, has the potential to outperform normal digital circuitry even though it is based on analog circuit elements. Still the most far reaching potential of it maybe the high dimensional multi-value logic type working modes provided their mathematical theory is available.

After introducing this system, there are more open questions than answered. Further efforts should address the realization, speed and power dissipation of practical circuit elements, including the actual analog circuits.

This paper dealt with the logic elements only. Control elements needed for synchronous or asynchronous processors have not been addressed. Similarly the possibility of using local stochastic controls utilizing local reference noises has not been dealt with. Thus the possibility of a standalone noise-based microprocessor remains the object of future studies.

## 14. Acknowledgements